\documentclass{article}
\usepackage{ijcai24}

\usepackage{times}
\usepackage{soul}
\usepackage{url}
\usepackage[hidelinks]{hyperref}
\usepackage[utf8]{inputenc}
\usepackage[small]{caption}
\usepackage{graphicx}
\usepackage{xcolor}
\usepackage{amsmath}
\usepackage{amsthm}
\usepackage{booktabs}
\usepackage{algorithm}
\usepackage{algorithmic}
\usepackage[switch]{lineno}

\usepackage{booktabs}
\usepackage{multirow}
\usepackage{tabularx}
\usepackage{subfigure}
\usepackage{enumitem}

\title{A Survey on Data-Centric Recommender Systems}

\author{
Riwei Lai$^1$\and
Rui Chen$^2$\And
Chi Zhang$^2$\\
\affiliations
$^1$Department of Computer Science, Hong Kong Baptist University\\
$^2$College of Computer Science and Technology, Harbin Engineering University\\
\emails
csrwlai@comp.hkbu.edu.hk,
\{ruichen, zhangchi20\}@hrbeu.edu.cn
}

\begin{document}

\maketitle

\begin{abstract}
Recommender systems (RSs) have become an essential tool for mitigating information overload in a range of real-world applications. Recent trends in RSs have revealed a major paradigm shift, moving the spotlight from model-centric innovations to data-centric efforts (e.g., improving data quality and quantity). This evolution has given rise to the concept of data-centric recommender systems (Data-Centric RSs), marking a significant development in the field. This survey provides the first systematic overview of Data-Centric RSs, covering 1) the foundational concepts of recommendation data and Data-Centric RSs; 2) three primary issues of recommendation data; 3) recent research developed to address these issues; and 4) several potential future directions of Data-Centric RSs.
\end{abstract}

\section{Introduction}
Recommender systems (RSs) have been widely adopted to alleviate information overload in various real-world applications, such as social media, e-commerce, and online advertising. The past few decades have witnessed the rapid development of recommendation models, evolving from traditional collaborative-filtering-based models~\cite{RFG09} to more advanced deep-learning-based ones~\cite{GRI23}, which have markedly improved the accuracy, diversity, and interpretability of recommendation results~\cite{LZL20}.

However, as recommendation models are growing larger and increasingly complex, as exemplified by the P5 recommendation model~\cite{GLF22} that integrates five recommendation-related tasks into a shared language generation framework, the primary constraint impacting recommendation performance gradually transitions towards recommendation data. Instead of focusing solely on developing even more advanced models, an increasing number of researchers have been advocating for the enhancement of recommendation data, leading to the emergence of the novel concept of data-centric recommender systems (Data-Centric RSs)~\cite{ZBL23}.

The fundamental rationale behind Data-Centric RSs is that data ultimately dictates the upper limits of model capabilities. Large and high-quality data constitutes the essential prerequisite for breakthroughs in performance. For instance, the remarkable advancements of GPT models in natural language processing are mainly originated from the use of huge and high-quality datasets~\cite{OWJ22}. Similarly, in computer vision, convolutional neural networks exhibit performance on par with vision transformers when they have access to web-scale datasets~\cite{SBB23}. For RSs, the implication is clear: the greater the quality~\cite{WFH21} and/or the quantity~\cite{LYZ23} of recommendation data, the more proficiently RSs can characterize user preferences, resulting in recommendations that resonate well with users.

Despite considerable attention from researchers and a great variety of methods that have been put forth in Data-Centric RSs, to the best of our knowledge, there has been no effort to gather and provide a summary of works in this promising and fast-developing research field. To fulfill this gap, we conduct a systematic review of the literature on Data-Centric RSs and provide insights from the following four aspects. 
We first detail the specifics of data that could be used for recommendation and present the formalization of Data-Centric RSs (\textbf{Section 2}). Next, we identify three key issues in recommendation data, i.e., data incompleteness, data noise, and data bias (\textbf{Section 3}), and categorize the existing literature in accordance with these issues (\textbf{Section 4}). Finally, we spotlight a number of encouraging research directions in Data-Centric RSs (\textbf{Section 5}).

\begin{figure}[t]
  \centering
  \includegraphics[width=0.85\linewidth]{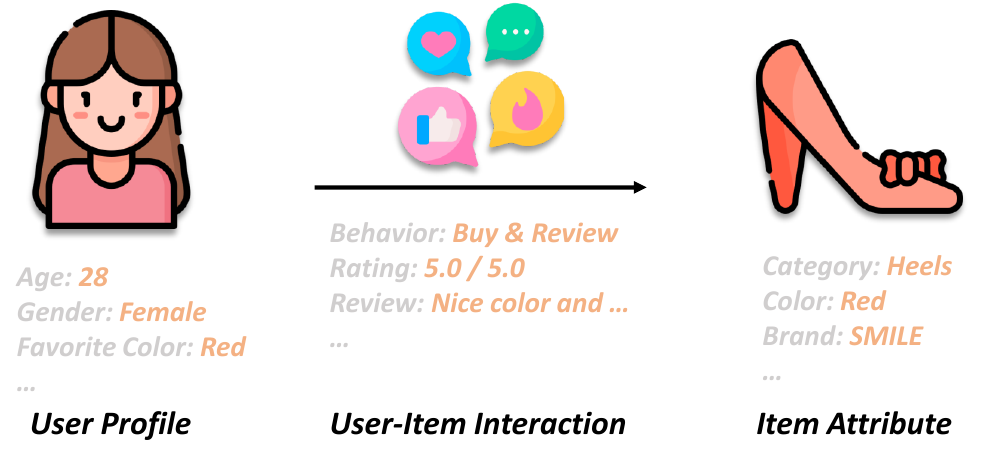}
  \caption{Three types of recommendation data.}
  \label{fig:data}
  \vspace{-3mm}
\end{figure}

\section{Formulation}
\subsection{Recommendation Data}
The goal of RSs is to help users discover potential items of interest and then generate personalized recommendations. To achieve this goal, as shown in Figure~\ref{fig:data}, we summarize three types of data that can be used in RSs:

\begin{itemize}[leftmargin=*]
    \item \textit{User profiles}: User profiles refer to a collection of information and characteristics that describe an individual user. In the context of recommendation, user profiles typically include demographics, preferences, behavior patterns, and other relevant data that helps define and understand a user's characteristics, needs, and interests.
    
    \item \textit{Item attributes}: Item attributes involve the specific details or characteristics that describe an item. These details can include color, material, brand, price, and other relevant information that helps identify or categorize the item.
    
    \item \textit{User-item interactions}: User-item interactions represent users' actions or involvements with items. Additional contextual information such as ratings, behaviors, timestamps, and reviews can also be utilized to provide a more comprehensive picture of the interactions between users and items.
\end{itemize}

\begin{figure}[t]
  \centering
  \includegraphics[width=\linewidth]{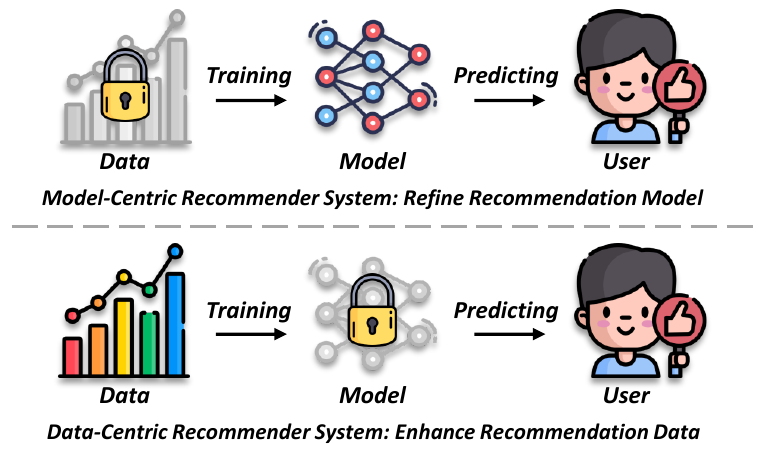}
  \caption{Model-Centric RSs v.s. Data-Centric RSs.}
  \label{fig:dcrs}
\end{figure}

\subsection{Data-Centric RSs}
As shown in Figure~\ref{fig:dcrs}, different from Model-Centric RSs which improve recommendation performance by refining models, Data-Centric RSs shift the focus from model to data and emphasize the importance of data enhancement. More specifically, given the recommendation data $\mathcal{D}$ (e.g., user-item interactions), Data-Centric RSs aim to determine a strategy $f$, which takes the original data $\mathcal{D}$ as input and outputs the enhanced data $\mathcal{D}^\prime$:
\begin{align}
    \mathcal{D}^\prime = f(\mathcal{D}).
    \label{eq:def}
\end{align}
The enhanced data $\mathcal{D}^\prime$ could be used by different recommendation models to further improve their performance. We also attempt to answer the following questions to clarify the definition of Data-Centric RSs:

\textit{Q1: Are Data-Centric RSs the same as Data-Driven RSs?} 
Data-Centric RSs and Data-Driven RSs are fundamentally different, as the latter only emphasize the use of recommendation data to guide the design of recommendation models without modifying the data, which are essentially still Model-Centric RSs~\cite{ZBL23}.

\textit{Q2: Why Data-Centric RSs are necessary?} 
The objective of a recommendation model is to fit the recommendation data. Without changing the data, Model-Centric RSs may generalize or even amplify errors (e.g., noise or bias) in the data, resulting in suboptimal recommendations~\cite{ZMZ23}. Therefore, Data-Centric RSs are necessary.

\textit{Q3: Will Data-Centric RSs substitute Model-Centric RSs?} 
Data-Centric RSs do not diminish the value of Model-Centric RSs. Instead, these two paradigms complement each other to build more effective recommender systems. Interestingly, model-centric methods can be used to achieve data-centric goals. For example, diffusion models~\cite{LYZ23} can be an effective tool for data augmentation. \textcolor{black}{Data-centric methods can facilitate the improvement of model-centric outcomes. For instance, high-quality synthesized data could inspire further advancements in model design~\cite{SDW22}}.

\begin{figure}[t]
  \centering
  \includegraphics[width=\linewidth]{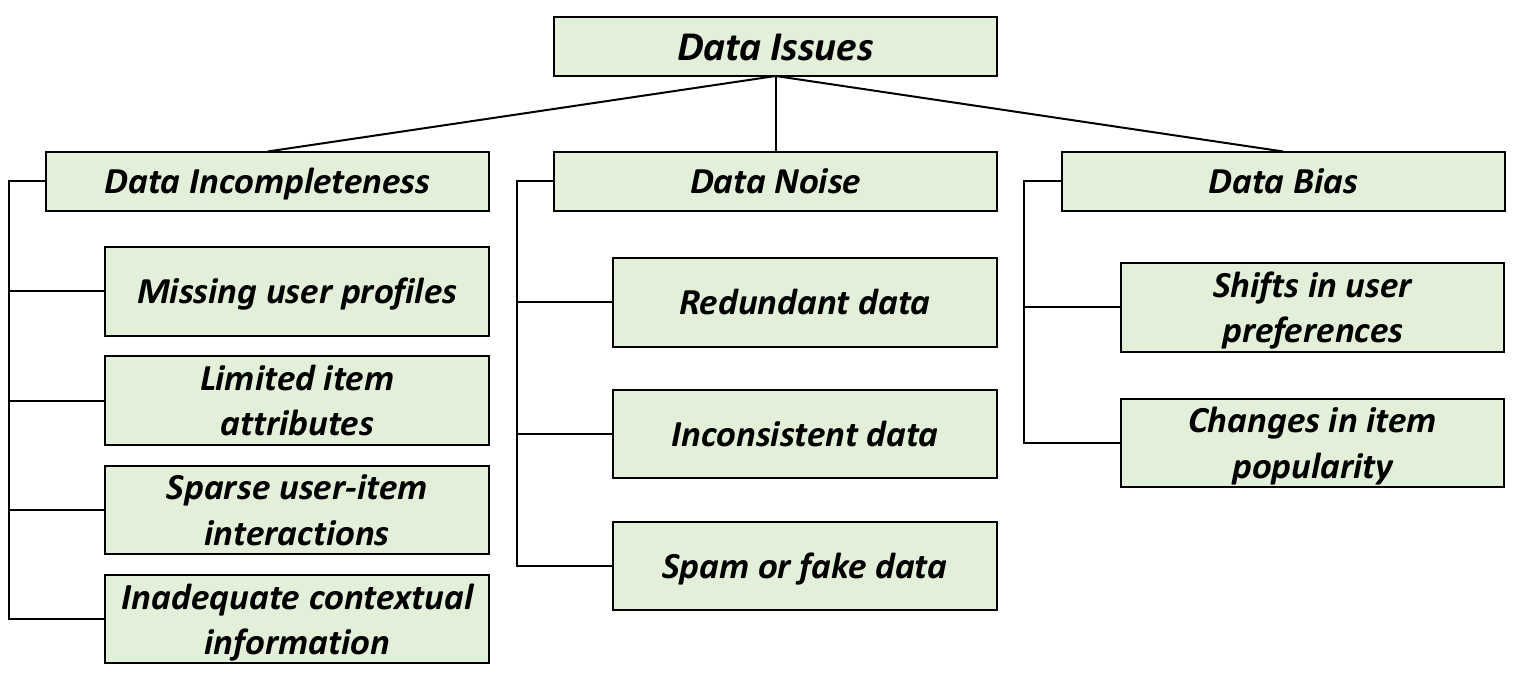}
  \caption{Overview of data issues in RSs.}
  \label{fig:dataiss}
\end{figure}

\section{Data Issues}
As illustrated in Figure~\ref{fig:dataiss}, we identify three key issues from which recommendation data may suffer, including \textbf{data incompleteness}, \textbf{data noise}, and \textbf{data bias}.

\subsection{Data Incompleteness}
\label{sec:datainc}
The data incompleteness issue refers to the scenarios where data used for making recommendations is inadequate, consequently resulting in information gaps or missing details regarding users or items. Specifically, the forms and reasons in which recommendation data can be incomplete include:
\begin{itemize}[leftmargin=*]
    \item \textit{Missing user profiles}: During the registration or setup process, users may fail to fully complete their profiles. Key information may be omitted as they might bypass certain fields or provide inadequate information~\cite{WRT23}. For example, a user may neglect to add age or gender, resulting in a profile that is less informative than it could be.
    
    \item \textit{Limited item attributes}: Similarly, data regarding item attributes may also be lacking. This incompleteness hinders the ability to precisely portray items and capture their distinct characteristics~\cite{WYZ20}. For instance, an online bookshop may only provide basic data about a book such as the title and author, neglecting additional details like genre or publication year that can enhance the recommendation accuracy.
    
    \item \textit{Sparse user-item interactions}: RSs can encounter the ``cold start'' problem when new users join. With limited or no historical interactions for these users, providing accurate recommendations becomes challenging~\cite{WYW19}. Additionally, users usually do not engage with every item. For example, in a movie recommender system, a user may only rate a handful of thousands of movies available, leading to an incomplete picture of true preferences.
    
    \item \textit{Inadequate contextual information}: Contextual information like timestamps, ratings, locations, or user reviews significantly contributes to generating effective recommendations. However, due to privacy issues or constraints in user feedback, the available contextual information may be incomplete or lack details~\cite{CKK19}. For instance, a user might give a high rating for a hotel visit but does not provide a review that could offer useful insights about his/her preferences for future recommendations.
\end{itemize}

\subsection{Data Noise}
\label{sec:datanoi}
Data noise arises when a portion of the recommendation data is irrelevant or erroneous, which negatively impacts the quality of recommendations provided. Noisy recommendation data can appear in several forms:
\begin{itemize}[leftmargin=*]
    \item \textit{Redundant data}: An abundance of identical data is often the consequence of errors during the data collection process. RSs might incorrectly identify and log certain identical item attributes multiple times, such as the same item category appearing repeatedly. Alternatively, it might record a user interacting with the same item multiple times -- a shop page refresh mistake, for example.
    
    \item \textit{Inconsistent data}: Inconsistencies in data occur mainly due to human errors, such as users unintentionally clicking or providing incorrect ratings to an item~\cite{LWC23}. Additionally, different data sources, like explicit ratings, implicit signals, and textual reviews, can present conflicting information about a user's current preferences. For example, a user might provide a low rating for a product, but the text in his/her review might be generally positive, which creates confusion.
    
    \item \textit{Spam or fake data}: At a more malicious level, RSs can be susceptible to spam or fake data generated by either malicious users or automated bots trying to harm the system. This undesired data can significantly contaminate the pool of authentic user data and can drastically impact the quality of recommendations, leading to decreased user satisfaction~\cite{WFL23}. A typical example is ``review bombing'', where orchestrated attempts by certain users or bots fill a product's review section with negative feedback to harm its overall rating, even though the product may be generally well-received by the majority.
\end{itemize}

\begin{figure}[t]
  \centering
  \includegraphics[width=\linewidth]{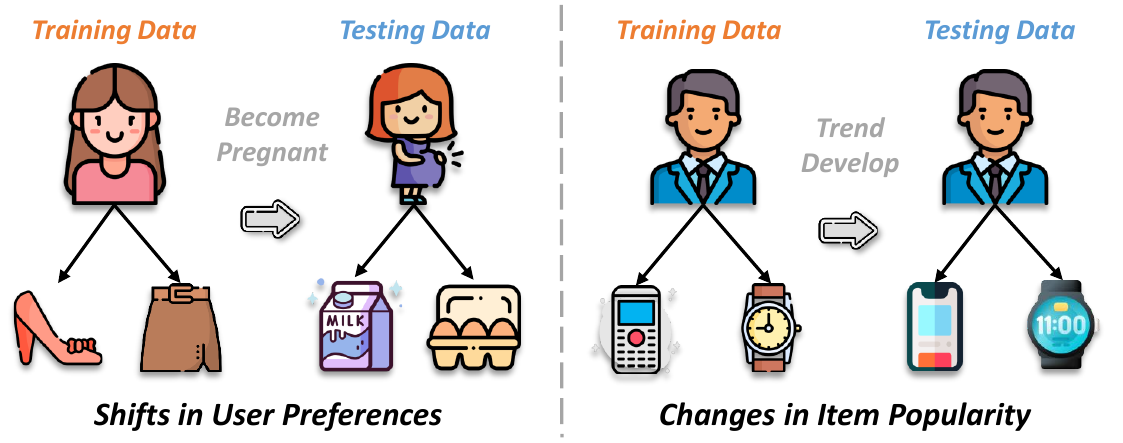}
  \caption{An illustration of data bias in RSs.}
  \label{fig:databias}
\end{figure}

\subsection{Data Bias}
\label{sec:databias}
A significant data bias issue arises when there is a significant distribution shift between the data collected and the real-world data. This problem mainly originates from:
\begin{itemize}[leftmargin=*]
    \item \textit{Shifts in user preferences}: As illustrated in Figure~\ref{fig:databias}, users' preferences can change due to shifts in wider environmental factors or their personal circumstances like pregnancy. In these scenarios, data collected in the past may no longer provide an accurate representation of the user's current preferences~\cite{WLW23}.

    \item \textit{Changes in item popularity}: Similarly, the popularity of certain items or categories is not static and can significantly vary over time. Items or trends that were once prevalent may lose their charm and relevance as time passes. For example, as shown in Figure~\ref{fig:databias}, a certain genre of products like the watch that was the rage a few years ago may not hold the same appeal to the audience in the present day as tastes evolve and new genres emerge~\cite{ZMZ23}.
\end{itemize}

\begin{table*}[t]
\centering
\begin{tabular}{c|l|l}
\hline
Addressed Issues & \multicolumn{1}{c|}{Categories} & \multicolumn{1}{c}{Representative Work and Key Technology}  \\ \hline
\multirow{6}{*}{Data Incompleteness}
& \multirow{3}{*}{Attribute Completion} & HGNN-AC~\cite{WYZ20}: Graph Neural Networks \\ 
&  & AutoAC~\cite{ZZWX23}: Neural Architecture Search \\ 
& & LLMRec~\cite{WRT23}: Large Language Models \\ \cline{2-3} 
& \multirow{3}{*}{Interaction Augmentation} & AugCF~\cite{WYW19}: Generative Adversarial Nets \\  
&  & DENS~\cite{LCZ23}: Disentanglement Learning \\ 
& & DiffuASR~\cite{LYZ23}: Diffusion Models \\ \hline
\multirow{7}{*}{Data Noise} 
& \multirow{3}{*}{Reweighting-Based Denoising} & R-CE~\cite{WFH21}: Adaptive Denoising Training \\
&  &  SGDL~\cite{GDH22}: Self-Guided Learning \\
&  &  AutoDenoise~\cite{GRI23}: Reinforcement Learning \\ \cline{2-3} 
& \multirow{3}{*}{Selection-Based Denoising} & T-CE~\cite{WFH21}: Adaptive Denoising Training \\
&  & STEAM~\cite{LWC23}: Self-Supervised Learning \\ 
&  & DeCA~\cite{WXM23}: Cross-Model Agreement \\\cline{2-3} 
& \multirow{1}{*}{Dataset Distillation/Condensation} &  DISTILL-CF~\cite{SDW22}: Dataset Meta-Learning \\ \hline
\multirow{6}{*}{Data Bias} 
& \multirow{3}{*}{User Preference Alignment} & Aspect-MF~\cite{ZMB19}: Latent Factor Models \\ 
& & DICE~\cite{ZGL21}: Disentanglement Learning \\
& & CDR~\cite{WLW23}: Variational AutoEncoder \\\cline{2-3}
& \multirow{3}{*}{Item Popularity Alignment} & IPS-MF~\cite{SYN20}: Inverse Propensity Scoring \\ 
& & MACR~\cite{WFC21}: Counterfactual Inference   \\
& & InvCF~\cite{ZZW23}: Invariant Learning \\ \hline
\end{tabular}
\caption{Representative works and key techniques used in handling different data issues.}
\label{tab:work}
\end{table*}

\section{Research Progress}
We organize the existing literature according to the data issues in RSs outlined before. Specific categorization as well as representative works and techniques are shown in Table~\ref{tab:work}.
\subsection{Handling Incomplete Data}
The key to handling incomplete data is to \textit{fill in} the missing information. According to the different forms of incomplete data introduced in Section~\ref{sec:datainc}, we divide existing methods into two categories: \textbf{attribute completion} and \textbf{interaction augmentation}.

\subsubsection{Attribute Completion}
Let $\mathcal{V}$ be the set of users and items. In practice, the profiles or attributes of some users or items may not be available. Therefore, the set $\mathcal{V}$ can be divided into two subsets, i.e., $\mathcal{V}^+$ and $\mathcal{V}^-$, which denote the attributed set and the no-attribute set, respectively. Let $\mathcal{A} = \{a_v \mid v \in \mathcal{V}^+\}$ denote the input attribute set. Attribute completion aims to complete the attribute for each no-attribute user or item $v \in \mathcal{V}^-$:
\begin{align}
    \mathcal{A}^C = \{a_v^c \mid v \in \mathcal{V}^-\} = f_{ac}(\mathcal{A}),
    \label{eq:ac}
\end{align}
where $a_v^c$ denotes the completed attribute of $v$. The enhanced input attribute set $\mathcal{A}^\prime$ is formulated as:
$$\mathcal{A}^\prime = \mathcal{A} \bigcup \mathcal{A}^C = \{a_v \mid v \in \mathcal{V}^+\} \bigcup \{a_v^c \mid v \in \mathcal{V}^-\}.$$

Existing works on attribute completion mainly rely on \textit{utilizing the topological structure of given graphs} (e.g., user-item interaction, knowledge graphs, and social networks)~\cite{WYZ20,YMD20,JHL21,TZL22,ZZWX23,GCL23}. For instance, by modeling user-item interactions with respective user or item attributes into an attributed user-item bipartite graph, AGCN~\cite{WYZ20} proposes an adaptive graph convolutional network for joint item recommendation and attribute completion, which iteratively performs two steps: graph embedding learning with previously learned attribute values, and attribute update procedure to update the input of graph embedding learning. Moreover, given a knowledge graph, HGNN-AC~\cite{JHL21} leverages the topological relationship between nodes as guidance and completes attributes of no-attribute nodes by weighted aggregation of the attributes of linked attributed nodes. AutoAC~\cite{ZZWX23} identifies that different attribute completion operations should be taken for different no-attribute nodes and models the attribute completion problem as an automated search problem for the optimal completion operation of each no-attribute node. Instead of focusing on attribute completion accuracy, FairAC~\cite{GCL23} pays attention to the unfairness issue caused by attributes and completes missing attributes fairly.

Given the extensive knowledge base and powerful reasoning capabilities of large language models (LLMs)~\cite{ZZL23}, some recent works have focused on \textit{leveraging LLMs to complete missing attributes}. For example, LLMRec~\cite{WRT23} carefully designs some prompts as the inputs of ChatGPT~\cite{OWJ22} to generate user profiles or item attributes that were not originally part of the dataset. An example of designed prompts is as follows:

\noindent \framebox[0.95\width]{
    \begin{tabular}{l}
        \textit{Provide the inquired information of the given movie.} \\
        \textit{\textbf{Heart and Souls (1993), Comedy/Fantasy}} \\
        \textit{The inquired information is: \textbf{director, country, language}.} \\
        \textit{Please output them in form of: director, country, language.}
    \end{tabular}
}
Similar to FairAC, some works~\cite{XWL23} are exploring the ability of LLMs in generating sensitive user profiles or item attributes and considering the risks it may bring to privacy leakage and unfairness issues.

\subsubsection{Interaction Augmentation}
Let $\mathcal{O} = \{(u, i) \mid u, i \in \mathcal{V}\}$ be the set of user-item pair $(u, i)$ with observed interactions. We denote $\mathcal{R} = \{r_{u,i} \mid (u, i) \in \mathcal{O}\}$ as the input interaction set. For inactive users and unpopular items, insufficient observed interactions can lead to inaccurate characterization of user preferences and item features. Therefore, interaction augmentation aims to augment the interaction information of some specific unobserved user-item pairs $(u,i) \notin \mathcal{O}$:
\begin{align}
    \mathcal{R}^A = \{r_{u,i}^a \mid (u,i) \notin \mathcal{O}\} = f_{ia}(\mathcal{R}),
    \label{eq:ia}
\end{align}
where $r_{u,i}^a$ denotes the augmented interaction information of $(u, i)$. The enhanced input interaction set $\mathcal{R}^\prime$ is:
$$\mathcal{R}^\prime = \mathcal{R} \bigcup \mathcal{R}^A = \{r_{u,i} \mid (u,i) \in \mathcal{O}\} \bigcup \{r_{u,i}^a \mid (u,i) \notin \mathcal{O}\}.$$

\begin{figure*}[t]
  \centering
  \includegraphics[width=\linewidth]{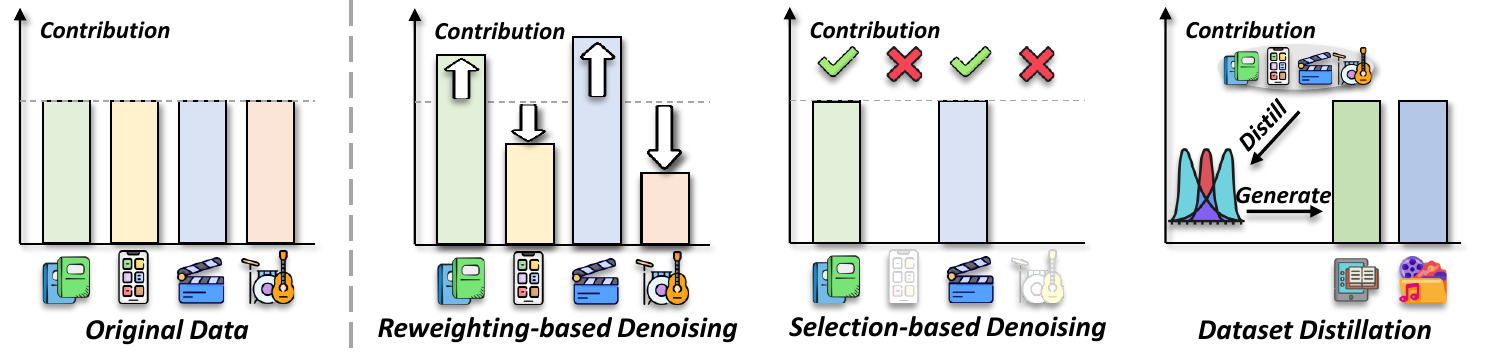}
  \caption{Categorization of data denoising methods in RSs.}
  \label{fig:datadeno}
\end{figure*}

For implicit information such as like/dislike, the critical focus of interaction augmentation lies in \textit{how to choose an interaction}. Negative sampling pays attention to \textit{how to choose an interaction as dislike}~\cite{RFG09,ZCW13,CSS17,DQY20,HDD21,LCZ23,SCF23,LCH24}. Specifically, negative sampling aims to identify uninteracted items of a user as negative items. The simplest and most prevalent idea is to randomly select uninteracted items, BPR~\cite{RFG09} is a well-known instantiation of this idea. Inspired by the word-frequency-based negative sampling distribution for network embedding~\cite{MSC13}, NNCF~\cite{CSS17} adopts an item-popularity-based sampling distribution to select more popular items as negative. DNS~\cite{ZCW13} proposes to select uninteracted items with higher prediction scores (e.g., the inner product of a user embedding and an item embedding). Such hard negative items can provide more informative training signals so that user interests can be better characterized. SRNS~\cite{DQY20} oversamples items with both high predicted scores and high variances during training to tackle the false negative problem. DENS~\cite{LCZ23} points out the importance of disentangling item factors in negative sampling and designs a factor-aware sampling strategy to identify the best negative item. MixGCF~\cite{HDD21} synthesizes harder negative items by mixing information from positive items while AHNS~\cite{LCH24} proposes to adaptively select negative items with different hardness levels.

Different from negative sampling, positive augmentation focuses on \textit{how to choose an interaction as like}~\cite{WYW19,WZX21,YWH21,ZYZ21,LXC22,LYZ23,WRT23}. For example, EGLN~\cite{YWH21} selects uninteracted items with higher prediction scores and labels them as positive to enrich users' interactions. CASR~\cite{WZX21} leverages counterfactual reasoning to generate user interaction sequences in the counterfactual world. MNR-GCF~\cite{LXC22} constructs heterogeneous information networks and fully exploits the contextual information therein to identify potential user-item interactions. Based on generative adversarial nets (GANs), AugCF~\cite{WYW19} generates high-quality augmented interactions that mimic the distribution of original interactions. Inspired by the superior performance of diffusion models in image generation, DiffuASR~\cite{LYZ23} adapts diffusion models to user interaction sequence generation, and a sequential U-Net is designed to capture the sequence information and predict the added noise of generated interactions. Leveraging the capabilities of LLMs, LLMRec~\cite{WRT23} also seeks to identify both positive and negative interactions from a candidate set using well-designed prompts.

For contextual information like rating, the critical focus of interaction augmentation shifts to \textit{how to infer the missing value}~\cite{RLZ12,RLZ13,YJS18,CKK19,LJM19,CKC20,HC22}. For instance, AutAI~\cite{RLZ12} and AdaM~\cite{RLZ13} calculate missing ratings according to heuristic similarity-based metrics, such as Pearson correlation coefficient or cosine similarity. RAGAN~\cite{CKK19} and UA-MI~\cite{HC22} leverage GANs for rating augmentation. Instead of augmenting ratings of interactions between real users and items, AR-CF~\cite{CKC20} proposes to generate virtual users and items and then adopts GANs to predict ratings between them.

\subsubsection{Discussion}
While a variety of methods exist for addressing the incomplete data issue, the fact remains that no single method is capable of comprehensively addressing all scenarios involving missing data. Consequently, a considerable amount of time and effort must be devoted to identifying missing information and selecting enhancement strategies. Furthermore, \textit{evaluating the quality of enhanced data} is not straightforward -- improvements in RSs might be misleadingly attributed to the simple expansion of data volume, which can cloud the actual effects of refinements in data quality.

\subsection{Handling Noisy Data}
The key to handling the data noise issue is to \textit{filter out} the noisy information. According to the varying severity of noisy data presented in Section~\ref{sec:datanoi}, we divide existing denoising methods into three categories: \textbf{reweighting-based methods}, \textbf{selection-based methods}, and \textbf{dataset distillation/condensation}, which are illustrated in Figure~\ref{fig:datadeno}.

\subsubsection{Reweighting-Based Denoising}
The reweighting-based method aims to \textit{assign lower weights to the noisy data} (or \textit{assign higher weights to the noiseless data})~\cite{WFH21,GDH22,ZYZ22,ZJK23,GRI23,WGL23}. Wang \textit{et al.}~\cite{WFH21} experimentally observe that noisy interactions are harder to fit in the early training stages, and, based on this observation, they regard the interactions with large loss values as noise and propose an adaptive denoising training strategy called R-CE, which assigns different weights to noisy interactions according to their loss values during training. SLED~\cite{ZJK23} identifies and determines the reliability of interactions based on their related structural patterns learned on multiple large-scale recommendation datasets. FMLP-Rec~\cite{ZYZ22} adopts learnable filters for denoising in sequential recommendation. The learnable filters perform a fast Fourier transform (FFT) to convert the input sequence into the frequency domain and filter out noise through an inverse FFT procedure. SGDL~\cite{GDH22} leverages the self-labeled memorized data as guidance to offer denoising signals without requiring any auxiliary information or defining any weighting functions. AutoDenoise~\cite{GRI23} adopts reinforcement learning to automatically and adaptively learn the most appropriate weight for each interaction.

\subsubsection{Selection-Based Denoising}
Instead of assigning lower weights, the selection-based method \textit{directly removes the noisy data}~\cite{TWL21,WFH21,YSS21,ZDZ22,LZW23,ZCZ23,QDG23,WXM23,LWC23}. For instance, different from R-CE, Wang \textit{et al.}~\cite{WFH21} propose another adaptive denoising training strategy called T-CE, which discards the interactions with large loss values. RAP~\cite{TWL21} formulates the denoising process as a Markov decision process and proposes to learn a policy network to select the appropriate action (i.e., removing or keeping an interaction) to maximize long-term rewards. DSAN~\cite{YSS21} suggests using the entmax function to automatically eliminate the weights of irrelevant interactions. HSD~\cite{ZDZ22} learns both user-level and sequence-level inconsistency signals to further identify inherent noisy interactions. DeCA~\cite{WXM23} finds that different models tend to make more consistent agreement predictions for noise-free interactions, and utilizes predictions from different models as the denoising signal. GDMSR~\cite{QDG23} designs a self-correcting curriculum learning mechanism and an adaptive denoising strategy to alleviate noise in social networks. STEAM~\cite{LWC23} further designs a corrector that can adaptively apply ``\textit{keep}'', ``\textit{delete}'', and ``\textit{insert}'' operations to correct an interaction sequence.

Some studies integrate the reweighting-based method with the selection-based method for better denoising~\cite{TXL22,YLY23}. For instance, RGCF~\cite{TXL22} and RocSE~\cite{YLY23} estimate the reliability of user-item interactions using normalized cosine similarity between their respective embeddings. Subsequently, they filter out interactions and only retain those whose weights exceed a pre-defined threshold value.

\subsubsection{Dataset Distillation/Condensation}
Dataset distillation or dataset condensation techniques~\cite{YLW23} aim to synthesize data points with the goal of condensing the comprehensive knowledge from the entire dataset into a small, synthetic data summary. This process retains the essence of the data, enabling models to be trained more efficiently. Recently, some works~\cite{SDW22,WFL23} observe that dataset distillation has a strong data denoising effect in recommendation. For example, DISTILL-CF~\cite{SDW22} applies dataset meta-learning to synthetic user-item interactions. Remarkably, models trained on the condensed dataset synthesized by DISTILL-CF have demonstrated improved performance compared to those trained on the full, original dataset.

\subsubsection{Discussion}
Normally, data collected from real-world scenarios are frequently contaminated with noise stemming from system bugs or user mistakes. However, obtaining labels for this noise is often impractical or even impossible, due to the lack of expert knowledge required for identifying noise, the high costs associated with manual labeling, or the dynamic nature of some noise which makes it hard to give fixed labels. In the absence of the ground truth, it is difficult to determine whether the denoising method achieves the optimal situation -- \textit{no over-denoising or under-denoising}. Taking into account this limitation, it may be preferable to synthesize a noise-free dataset via dataset distillation/condensation rather than attempting to adjust or filter the existing dataset through reweighting-based or selection-based methods.

\subsection{Handling Biased Data}
The key to handling biased data is to \textit{align} the biased training distribution with the unbiased test distribution. According to the causes of biased data explained in Section~\ref{sec:datanoi}, we divide existing debiasing methods into two categories: \textbf{user preference alignment} and \textbf{item popularity alignment}.

\subsubsection{User Preference Alignment}
User preferences may shift due to a variety of reasons, such as \textit{temporal changes}~\cite{ZMB19,WW21,DLH22}, \textit{locational moves}~\cite{YZC16}, or \textit{alterations in personal and environmental conditions}~\cite{ZGL21,HWC22,WLW23}. Existing methods are designed to track and adjust to these changes, thereby maintaining alignment with the ever-evolving user preferences. For example, Aspect-MF~\cite{ZMB19} analyzes and models temporal preference drifts using a component-based factorized latent approach. MTUPD~\cite{WW21} adopts a forgetting curve function to calculate the correlations of user preferences across time. ST-LDA~\cite{YZC16} learns region-dependent personal interests and crowd preferences to align locational preference drifts. Wang \textit{et al.}~\cite{WLW23} review user preference shifts across environments from a causal perspective and inspect the underlying causal relations through causal graphs. Based on the causal relations, they further propose the CDR framework, which adopts a temporal variational autoencoder to capture preference shifts.

\subsubsection{Item Popularity Alignment}
Existing methods for item popularity alignment roughly fall into five groups~\cite{ZMZ23}. \textit{Inverse propensity scoring}~\cite{SSS16,SYN20} utilizes the inverse of item popularity as a propensity score to rebalance the loss for each user-item interaction. \textit{Domain adaptation}~\cite{BV18,CXL20} leverages a small sample of unbiased data as the target domain to guide the training process on the larger but biased data in the source domain. \textit{Causal estimation}~\cite{WFC21,ZFH21,WLF22} identifies the effect of popularity bias through assumed causal graphs and mitigates its impact on predictions. \textit{Regularization-based methods}~\cite{BFM21} explore regularization strategies to adjust recommendation results, aligning them more closely with the actual popularity distribution. \textit{Generalization methods}~\cite{ZMW22,ZZW23,ZMZ23} aim to learn invariant features that counteract popularity bias, thereby enhancing the stability and generalization capabilities of recommendation models.

\subsubsection{Discussion}
Traditional evaluation settings in RSs may not be appropriate for assessing debiasing methods because they typically entail that the distribution of a test set is representative of the distribution in the training set (independent and identically distributed evaluation settings). Therefore, in this part, we discuss two \textit{out-of-distribution evaluation settings} to assess debiasing methods:
 
\begin{itemize}[leftmargin=*]
    \item \textit{Temporal split setting}: Temporal split setting slices the historical interactions into the training, validation, and test sets according to the timestamps~\cite{ZMW22,ZZW23}. In this case, any shift in user preferences or item popularity over time is appropriately represented and accounted for during the evaluation.
    \item \textit{Popularity split setting}: Popularity split setting constructs the training, validation, and test sets based on various popularity distributions~\cite{WFC21,ZGL21}. For example, the training interactions are sampled to be a long-tail distribution over items while the validation and test interactions are sampled with equal probability in terms of items (uniform popularity distribution). However, such a split setting has an inherent drawback: it may inadvertently lead to information leakage. By explicitly tailoring the test set to a known distribution of item popularity, the debiasing methods might be unduly influenced by this information.
\end{itemize}

\section{Future Directions}
\subsection{Data-Centric RSs with Multimodal Data}
Multimodal data refers to data that consists of multiple modalities or types of information, such as text, images, audio, video, or any combination thereof. Traditionally, RSs have primarily relied on user-item interaction data, such as ratings or click-through data, to generate recommendations. By incorporating multimodal data, RSs can capture richer and more diverse user preferences and item characteristics, leading to more personalized and relevant recommendations~\cite{TL19}. However, the data issues mentioned before (i.e., incomplete data, noisy data, and biased data) also exist in multimodal data, and they can pose additional challenges in the context of multimodal RSs:

\begin{itemize}[leftmargin=*]
    \item \textit{Heterogeneity}: Multimodal data can be highly heterogeneous, with different modalities having distinct data formats, scales, and distributions. For example, text data may require natural language processing techniques, while image data may need computer vision algorithms.
    \item \textit{Imbalance}: Multimodal datasets may exhibit imbalances in the distributions of different modalities. For example, there may be a larger number of text samples compared to images or audio samples. Modality imbalance can affect the performance and generalization of recommendation models trained on such data.
    \item \textit{Scalability}: Multimodal data, especially when it includes high-dimensional modalities like images or videos, can be computationally expensive to process and analyze. Therefore, handling large-scale multimodal data may require efficient algorithms or distributed computing frameworks to ensure scalability.
\end{itemize}

\subsection{Data-Centric RSs with LLMs}
With the emergence of large language models (LLMs) in natural language processing, there has been a growing interest in harnessing the power of these models to enhance recommender systems. In Data-Centric RSs, LLMs can serve as:
\begin{itemize}[leftmargin=*]
    \item \textit{Recommendation models}: Pre-trained LLMs can take as input a sequence that includes user profiles, item attributes, user-item interactions, and task instructions. Then LLMs analyze this information to understand the context and the user's preferences. Based on this understanding, LLMs can generate a sequence that directly represents the recommendation results, which could be a list of items, a ranking of items, or even detailed descriptions or reasons for the recommendations. However, using LLMs as recommendation models also raises some challenges such as limited token length or latency, especially for users with a large amount of interactions. With data denoising techniques to improve the design of input sequences, the ability of LLMs as recommendation models can be further explored.
    \item \textit{Data processors}: As mentioned before, given the extensive knowledge base and powerful reasoning capabilities of LLMs, some recent work has attempted to augment data with LLMs, for example, through carefully designed prompts, LLMRec~\cite{WRT23} employs three simple yet effective LLM-based data augmentation strategies to augment implicit feedback, user profiles, and item attributes, respectively. Moving forward, it's crucial to investigate the capability of LLMs in managing tasks such as data denoising and data debiasing. This could pave the way for LLMs to harmonize Data-Centric RSs effectively.
\end{itemize}

\subsection{Automatic Data-Centric RSs}
Automatic machine learning (AutoML)~\cite{HZC21} refers to the process of automating the end-to-end process of applying machine learning to real-world problems, which typically involves automating a variety of tasks that are part of the machine learning workflow. In the context of Data-Centric RSs, these tasks encompass data augmentation, data denoising, and data debiasing. Consequently, AutoML can automatically streamline and enhance the efficiency of these tasks, enabling more accurate recommendations, which is of great significance in practice.

\subsection{Transparent Data-Centric RSs}
Transparent Data-Centric RSs refer to Data-Centric RSs that not only offer enhanced data for model training but also provide insights into how and why particular enhancements are made, thereby allowing users and developers to understand the underlying decision-making processes. Research in transparent Data-Centric RSs is tackling complex challenges, such as the trade-off between transparency and complexity, ensuring user privacy while providing explanations, and developing standards for explainability and interpretability.

\section{Conclusions}
In this survey, we presented a comprehensive literature review of Data-Centric RSs. We systematically analyzed three critical issues inherent in recommendation data and subsequently categorized existing works in accordance with their focus on these issues. Additionally, we point out a range of prospective research directions to advance Data-Centric RSs. We expect that this survey can help readers easily grasp the big picture of this emerging field and equip them with the basic techniques and valuable future research ideas.

\clearpage
\bibliographystyle{named}
\bibliography{ijcai24}

\end{document}